\documentclass[twocolumn,showpacs,preprintnumbers,amsmath,amssymb]{revtex4}

\newcommand{\sect}[1]{\setcounter{equation}{0}\section{#1}}

\begin{document}

\title{{Do Rotations Beyond the Cosmological Horizon
\\Affect the Local  Inertial Frame?}}

\author{Ji\v ri Bi\v c\'ak,$^{1,2,4}$
Donald Lynden-Bell,$^{2,1,4}$ and Joseph Katz$^{3,1,4}$
\\$^1${\it Institute of Theoretical Physics,  Charles University, \it V Hole\v sovick\' ach 2, 180 00 Prague 8, Czech Republic}
\\{\it $^2$Institute of Astronomy, Madingley Road, Cambridge CB3 0HA,
United Kingdom}
 \\ {\it $^3$The Racah Institute of Physics, Givat Ram, 91904 Jerusalem,
Israel}
\\ {\it $^4$Max-Planck Institute for Gravitational Physics (Albert Einstein
Institute) \it 14476 Golm, Germany.}}


\begin{abstract}

If perturbations beyond the horizon have the velocities prescribed
everywhere then the dragging of inertial frames near the origin is
suppressed by an exponential factor. However if perturbations are
prescribed in terms of their angular momenta there is no such suppression.

We resolve this paradox and in doing so give new explicit results on the
dragging of inertial frames in closed, flat and open universe with and
without a cosmological constant.

\end{abstract}

\maketitle

\sect{Introduction}

 In a clearly written paper Schmid
\cite{Sc} describes how rotational perturbations of a spatially flat universe
influence the inertial frames.         He finds that such influences are
attenuated by an exponential Yukawa factor whenever the perturbation
lies beyond a `horizon'. He expressed his results in terms of a quantity
that he calls the energy current ${\vec J}_\epsilon$. What corresponds to
Schmid's energy current
${\vec J}_\epsilon$ in our calculation is a quantity ${\vec J}_s$ with the
dimensions of angular momentum. However his result appears to disagree
with our earlier result \cite{LKB95} (hereafter LKB) that in a flat universe
the rotation of the inertial frame \unskip, $\omega$, due to any system of spheres with small
rotations about a center is given by
\begin{equation} \vec\omega (\overline{r},t)=\frac {2G}{c^2}\left[
\frac{1}{\overline{r}{}^3}{\vec J}(t,<\overline{r})+\int_{\overline r}^{\infty}\frac{1}{\overline
r{'}^3}\frac{{\rm d}{\vec J}}{{\rm d}\overline
r{'}}\,{\rm d}\overline r{'}\right],
\label{1}
\end{equation}
where ${\vec J}(t,<\overline{r})$ is the total angular momentum within the sphere
of proper radius \cite{foot1} $\overline r=a(t)r$.
This expression demonstrates how angular momenta  at all distances contribute and
shows no exponential cut-off and no influence of any horizon. 

Both results agree, however,  that inertial `influences' may be expressed
instantaneously i.e. with no light travel-time delay. This is because they
follow from the constraint equations of General Relativity with an
appropriate mapping onto an unperturbed universe to provide a suitable
gauge.

The results are in {\it apparent} contradiction. However, when the details
of both calculations are examined it is evident that the contradiction lies
in the attribution of different causes for the effect. Schmid's `energy
current' is considered by him as the source of the rotational dragging of
inertial frames $\omega$. Schmid's quantity does not obey a conservation
law but for a rotating sphere it can be directly expressed in terms of its
angular velocity $\Omega$, density $\rho$, pressure $p$ and proper radius
$\overline r$ as
\begin{equation}
J_\epsilon =2\pi\!\!\int\!\!\!\int(\rho+p)\Omega\sin^3\theta~\overline r^4 \,{\rm d}\theta
\,{\rm d}\overline{r},
\end{equation}
$\Omega$ is the `coordinate' angular velocity, not that measured relative to
the local inertial frame which, as we show below, is the quantity needed
in the angular momentum conservation law. The contribution to the conserved angular momentum is
\begin{equation}
2\pi(\rho+p)(\Omega -\omega)\sin^3{\theta} \, \overline r^4 \,{\rm d}\theta \,{\rm d}\overline{r}.
\end{equation}
The difference comes about mathematically because the perturbed metric
is not diagonal but in spherical polar coordinates is
\begin{equation}
{\rm d}s^2={\rm d}t^2-a^2(t) \Big\{   {\rm d}r^2+r^2\Big[
{\rm d}\theta^2+\sin^2\theta \, ({\rm d}\varphi^2-2\omega \, {\rm d}\varphi \,{\rm d}t) 
\Big]\Big\},
\label{metric}
\end{equation}
where $\omega(t,r)$ gives the small rotation of the inertial frame  due to the rotational
perturbations everywhere.

For a perturbed 3-flat universe, angular momentum  conservation is given by the
equation 
\begin{equation}
\frac{\partial}{\partial x^\mu}\left(-\sqrt{-g}\,T^\mu_\nu\eta^\nu \right)=0,
\end{equation}
where $\eta^\nu$ is the angular Killing vector of the background (flat)
space corresponding to the particular component of angular momentum considered. 

Thus the conserved quantity is (the minus sign comes from the signature
we use in the metric)
\begin{equation}
J=\int-T^0_\varphi
\sqrt{-g}\,{\rm d}^3\!x.
\end{equation}
Since the $\varphi$ component of $T^{\mu\nu}$ is brought down in this
expression, it is {\it not} merely the motion of the fluid that is involved
in $T^0_\varphi$ via its contribution $u^\mu u^\nu$ but also the off-diagonal metric
component $g_{0\varphi}$ which depends on $\omega$ 
at the position of the source [see our metric (\ref{metric})]. As we see
from equation  (\ref{1}), we regard the conserved angular momentum  $J$ as the source
of the dragging of inertial frames, and this was the quantity we used in
LKB.

Schmid's work for a spatially flat universe is more general than our
work published so far, since he considers {\it all\/} vector perturbations,
nevertheless we treated closed and open universes as well as flat ones and
indeed from a Machian viewpoint it is the closed universes that are more
interesting by far. We also considered all spherical but inhomogeneous
Lema\^\i tre-Tolman-Bondi universes with rotational perturbations that
were constant on spheres. Finally we looked into the problem of the
rotation of inertial frames  induced by spheres of given angular velocities, rather
than given angular momentum \unskip. This is a special case of Schmid's problem  but
generalized to closed and open universes. In our discussion we wrote
down the equations governing $\omega(t,r)$ when $\Omega(t,r)$ was given
and showed how they could be solved. We carried out the detailed
solution only for the static closed Einstein universe (LKB Appendix~A).

Schmid's beautiful result that the dragging is exponentially suppressed
when a sphere of given angular velocity is outside his horizon, has
stimulated us to work out all our solutions in detail for all FRW universes.
Barring factors of $a(t)$ that Schmid seems to have omitted in error, we
fully confirm his result for a flat universe. Thus we have the fascinating
paradox that {\it while spheres of given angular velocity have their
dragging exponentially suppressed if they are outside the horizon,
nevertheless the dragging of spheres of given angular momentum  ~suffers no such
suppression!} How can this be!

In the prescribed angular momentum  problem one may consider (for an open or 3-flat
universe) having only one spherical shell of finite thickness having angular
momentum. The gravity of this source will induce a rotation of inertial frames 
everywhere. The fluid at all other places will respond inertially and start to
rotate so that
$\Omega=\omega$ everywhere except on the original shell. Thus in the
prescribed angular momentum  problem we give one shell angular momentum \unskip, sit back and watch. We
see how the inertial frames  are affected everywhere else merely by watching the
rotations of all other spheres.

The prescribed angular velocity problem needs more organization in the
creation of the initial state. If we start one thick shell rotating at the
prescribed rate then all the other will start moving so as to keep up with
the induced rotation of the inertial frames. If the prescription is to have just the one
thick shell rotating and none of the others we shall have to stop them. In
doing so we have to give them negative angular momentum  to keep $\Omega$
 zero even though the inertial frame  is rotating at $\omega$. When in Schmid's
problem the perturbation in angular velocity is considered as confined
beyond his horizon he shows that the rotation of inertial frames  ~is exponentially
suppressed near the origin. The prime reason is that in order to keep the
motion confined, the intervening spheres have to be given backward angular momentum 
to stop them from following their inertial tendency of rotating at the
inertial rate $\omega$. The influence of all the backward angular momentum  (of
non-rotating spheres!) rather effectively cancels most of the rotation of the
inertial frames  induced by the original shell. Hence the suppression of the effect is
due to all the negative angular momentum  ~that was supplied to keep the other  spheres
from rotating! The remaining suppression is due to the rotation of the
inertial frame  at the original shell itself: $\omega$ there is a fraction of $\Omega$ so that
$\Omega-\omega$, on which the source depends, is less.

There is a long history of treating dragging effects within spheres starting
with Einstein's treatment using an early version of his
gravitational theories. Within General Relativity the early works of
Thirring \cite{Th} and Lense and Thirring \cite{LT} were later generalized
to deep potential wells by Brill and Cohen \cite{BC}. This raised questions
as to whether the dragging would be perfect within a black-hole's horizon.
We believe that the first paper to remark on the apparent instantaneity if
inertial frames   is the pioneering paper of Lindblom and Brill \cite{LB} on
inertia in a sphere that falls through its horizon. More recently we explored
observational effects seen within such a sphere \cite{KLB98} and gave an
example of strong linear dragging in a rapidly accelerated charged
sphere \cite{LKB98}. Strong cosmological perturbations    in a weakly rotating
sphere surrounding a void were treated by Klein \cite {Kl}, and in greater
detail by Dole\v zel, Bi\v c\' ak, and Deruelle \cite{DBD} who also
discussed how an observer within such a cosmological  ~shell views the world
outside.

We owe a debt to Schmid as his work stimulated us to work out the
consequences of our solutions \cite{LKB95} in much greater detail and,
without that, we would never have raised, let alone understood, the
delightful paradox emphasized above. In particular we have now
investigated thoroughly the problem when $\omega$ is to be solved for with
the angular velocities given everywhere at one cosmic time. Previously we
had concentrated on the problem with the angular momenta  given. While both are
important problems we strongly believe that it is the latter that is of
dynamical importance in formulating Mach's Principle. It can
nevertheless be argued that the apparent agreement between the angular
positions of quasars at different epochs and the inertial frame  defined by using the
solar system as a giant gyroscope stimulates Machian ideas. While it is
the angular momentum  that is important for the physics it is the apparent kinematical
agreement between the {\it angular} velocity of the sphere and the inertial frame 
that is observed. In this sense the problem with given angular velocities
may be closer to Mach's original and it is unclear how distant
observations could measure the true angular momentum  of a sphere including its
dragging term, while its angular velocity is more directly observable.
However, see \cite{DBD} for the complications of light bending.
Unfortunately  the problem of the observed agreement of frames is not
that either Schmid or we have addressed since both our treatments
relate instantaneous quantities at the same cosmic time whereas
observers use no such world map (except in the solar system) but a world
picture in which distant objects are seen as they were long ago on the
backward light cone. It seems unlikely that an {\it exact} causal
relationship exists between proper motions of masses on our past light
cone and our local inertial frame, since, {\it at any cosmic time the inertial
frame's rotation has contributions from objects that were never in our past
light cone}. Of course such objects will no doubt have been seen by some
alien and the Copernican principle would suggest that the apparent
agreement of the kinematic and inertial frames  here will be repeated there. What is
under discussion above is the influence of distant bodies on the local
inertial frame. This is quite distinct from a comparison of the dynamics of
the solar system with its kinematics relative to distant quasars (as seen on a
hundred years of past light cones), from which the rotation of the inertial frame  is
computed.

Beside the resolution of the apparent contradiction with Schmid the main
contributions of this paper are the following.

{\it Section 2}. 
The derivation of the equations governing general perturbations  and a brief
introduction to Machian gauge conditions which allow the separation
of  the $(h_{0k})$  vector perturbations  equations from the others. The
discussion of the equations of motion that must be obeyed if the
contracted Bianchi identities are to be satisfied. As a consequence when
axial symmetry is imposed each ring of fluid preserves its angular momentum \unskip. This
section concludes with basic equations for odd parity axially symmetrical
perturbations  from which the remainder of the paper is derived.

{\it Section 3\/}
derives the explicit expressions for rotation of inertial frames  in terms of the angular momentum 
distribution at any one time. This is done for all FRW universes with
$k=\pm 1$ or $0$ but the simplest case is solved in this section
with $\Omega$ constant on  spheres at the time considered. This
corresponds to odd-parity  vector $l=1$ perturbations  with $\Omega$ independent
of $\theta$. In the following paper \cite{BLK2} (Paper~II) we allow for general  $\theta$ dependence. With the
integrals evaluated at fixed cosmic time and with the constants $c$ and
$G$ restored we have the following results for $\vec \omega(r)$ at fixed time
(for the derivation of the vector forms below see \cite{LKB95}): 

For $k=0$, $\overline r =a(t)r$,
\begin{equation}
\qquad \vec \omega=\frac{2G}{c^2a^3}\left[ 
{\vec J}(<r)r^{-3}+\int_{\overline r}^{\infty}\frac{{\rm d}{\vec J}}{{\rm d}
r} r^{-3}\,{\rm d} r\right]. 
\end{equation}
Notice that $\vec \omega\propto[a(t)]^{-3}$ since ${\vec J}$ is conserved.

For $k=1$,
\begin{equation}
\vec \omega=\frac{2G}{c^2a^3}\left[ 
{\vec J}(<\chi)W(\chi)+\int_{\chi}^{\pi}\frac{{\rm d}{\vec J}}{{\rm d}
\chi'}W(\chi')\,{\rm d} \chi'\right] + \vec \omega_0(t),
\end{equation}
here $\vec \omega_0(t)$ is undetermined, $r=\sin\chi$ and $W(\chi)=
\cot^3\chi+3\cot\chi$. The arbitrariness of  $\vec \omega_0(t)$ is intimately
connected with Mach's principle. The physical ${\vec J}$ involves
$(\vec \Omega-\vec \omega)$ and does not change for rotating axes as it involves a
difference, see \cite{LKB95} and below.

For $k=-1$,
\begin{equation}
\vec \omega=\frac{2G}{c^2a^3}\left[ 
{\vec J}(<\chi)\overline W(\chi)+\int_{\chi}^{\infty}\frac{{\rm d}{\vec J}}{{\rm d}
\chi'}\overline W(\chi')\,{\rm d}\chi'\right], 
\end{equation}
where $\overline W(\chi)=
\coth^3\chi-3\coth\chi+2$,
and $\overline W$ has an extra 2 so it tends to zero at $\chi\rightarrow\infty$ 
thus  ensuring that the boundary condition $\omega\rightarrow 0$ is
obeyed. When contributions from a $\theta$ dependence of $\Omega$ are
included these results are supplemented by  $\theta$  dependent terms
that average to zero on spheres. More general results are given in the accompanying
Paper~II.

{\it Section 4\/}
gives explicit solutions for the rotations of inertial frames  for the same special forms
of perturbations  as in Section 3 but now it is the angular velocities of the
different spheres that are given rather than their angular momenta  (this is {\it closer}
to what might be observed but cf. earlier discussion). We define $\lambda$ by
\begin{equation}
\lambda^2=2\kappa a^2(\rho+p)= 4(k-a^2\dot
H),
\label{la}
\end{equation}
$\kappa=8\pi G/c^4$, $\kappa=8\pi$ in geometrical units used in the following,
the dot denotes $\partial/\partial t$, $H=\dot a/a$ is the Hubble constant. 
The second relation
in (\ref{la})  follows from the combination of the background Einstein's
equations  for any $\rho$, $p$, $k$ and also for any value of the cosmological  constant
$\Lambda$. The rotation of inertial frames  near the origin due to an $\Omega$
distribution at large
$z'=\lambda r$ is for $k=0$
\begin{equation}
 \omega(r)=\frac{1}{3}\left(1+\frac{1}{10}\lambda^2r^2\right)
\int_0^\infty z'^2e^{-z'}\Omega(z')\,{\rm d}z',
\end{equation}
{\noindent which shows Schmid's exponential attenuation $e^{-z'}$. At the perturbation 
itself, close to $z_0$, we find for $z'$ large:}
\begin{equation}
\omega(r_0)=\frac{1}{2}\int_0^\infty
\left(\frac{z'}{z_0}\right)^2e^{-|z'-z_0|}\Omega(z')\,{\rm d}z'.
\end{equation}

For $k=1$ we give the results near the origin and at the perturbation  when
$\lambda^2>4$. When $\lambda^2<4$, which can occur when a $\Lambda$-term is
present, there is no exponential in the expression. It is assumed that
$\exp(\sqrt{\lambda^2-4}\,\chi)$ is large at the source. With $r=\sin\chi$
\begin{eqnarray}
\omega(\chi)&&=\frac{1}{3}\left(1+\frac{\lambda^2\chi^2}{10}\right)
\times\\\nonumber
&&\int_0^\infty\lambda^2\sqrt{\lambda^2-4}~e^{-\sqrt{\lambda^2-4}\chi'}\sin^2(\chi')
\Omega(\chi')\,{\rm d}\chi'.
\end{eqnarray}
We have assumed $\exp(\sqrt{\lambda^2-4}~x)\gg1$ for $x=\pi$, $\chi'$, 
and $\pi-\chi'$. At the `source' 
\begin{eqnarray}
&&\omega(\chi_0)=\\\nonumber
&&\frac{1}{2}  \int_0^\infty\!\!\lambda^2\frac{\lambda^2-3}{\sqrt{\lambda^2-4}}\left( 
\frac{\sin\chi'}{\sin\chi_0}\right)^2e^{-\sqrt{\lambda^2-4}|\chi'-\chi_0|}
\Omega(\chi')\,{\rm d}\chi'.
\end{eqnarray} 
Similarly for $k=-1$: $r=\sinh\chi$,
\begin{eqnarray}
&&\omega(\chi)=\frac{1}{3}\left(1+\frac{\lambda^2\chi^2}{10}\right)\times\\\nonumber
&&\int_0^\infty\lambda^2\sqrt{\lambda^2+4}~e^{-\sqrt{\lambda^2+4}\chi'}\sinh^2(\chi')
\Omega(\chi')\,{\rm d}\chi',
\end{eqnarray}
and at the source 
\begin{eqnarray}
&&\omega(\chi_0)=\\\nonumber
&&\frac{1}{2}  \int_0^\infty\frac{\lambda}{\sqrt{\lambda^2+4}}\left( 
\frac{\sinh\chi'}{\sinh\chi_0}\right)^2e^{-\sqrt{\lambda^2+4}|\chi'-\chi_0|}
\Omega(\chi')\,{\rm d}\chi'.
\end{eqnarray} 

We emphasize that all of the above relationships are true at any given
instant, but that both the angular momentum  distribution and the angular velocity  distribution at
later instants are related to those at earlier times, so can not be given
 {\it independently} of those given at an earlier epoch. In axial symmetry
the
angular momentum  distribution follows the motion of the perfect fluid but, as the angular momentum 
is first  order and the movement across the background is of first order,
the product can be neglected.
Thus to first order the angular momentum  density can be considered as painted on the
background. This   is not true of ${\vec J}_s$ which is not conserved and nor is
it true of the angular velocity  $\Omega$. In both cases to find  the time evolution one
must appeal to the equations of motion which, in axial symmetry, leads
back to local conservation of angular momentum density. Only by use of
its conservation can one find how $\Omega$ and ${\vec J}_s$ can evolve
consistently with Einstein equations (i.e. with the contracted Bianchi
identities). In this sense the given angular momentum  problem is far more physical than
either Schmid's problem or the given $\Omega$ problem to which it is
equivalent. The time evolution of $\omega$ and $\Omega$ are derived and
discussed in Section 5.

In a paper that has long been in gestation  we give a discussion of those
gauges in which the Machian relations of the local inertial frames  to the motions of
distant masses can be expressed instantaneously at constant cosmic time.
In that paper we derive all equations that govern all perturbations. All can be
solved using harmonics in the 3-space of constant time. However
harmonics are not as informative as Green's functions so in the following
paper \cite{BLK2} we
integrate the relationships between the rotations of the inertial frames  and the
angular momentum  density for all axially symmetrical odd-parity vector perturbations, called
usually "toroidal" perturbations in astrophysical and geophysical literature. 
These results allow $\Omega-\omega$, which enter the angular momentum   density, to be any
function of $(r,\theta)$ but independent of
$\varphi$. However, since the background is spherically symmetric, non-axisymmetric
perturbations can be generated by re-expanding axisymmetric perturbations
around a new axis, and taking the component with the new $e^{im\varphi}$
as the component with general $m$.

\sect{The equations to be solved}

We write the perturbed FRW metric in the form
\begin{eqnarray}
{\rm d}s^2&=&(\overline g_{\mu\nu}+h_{\mu\nu}){\rm d}x^\mu {\rm d}x^\nu
\\\nonumber
          &=& {\rm d}t^2-a^2(t)
f_{ij}{\rm d}x^i {\rm d}x^j+h_{\mu\nu}{\rm d}x^\mu {\rm d}x^\nu,
\label {ds2}
\end{eqnarray}
where the background  metric $\overline g_{\mu\nu}$ is used to move indices and the
time-independent part of the  spatial background metric
$f_{ij}~[i,j,k=1,2,3]$ is used to define the 3-covariant derivative $\nabla_k$ and
$\nabla^k=f^{kl}\nabla_l$.

In one of the standard coordinate systems  the background  FRW metric reads
\begin{equation}
{\rm d}s^2={\rm d}t^2-a^2\left[  \frac{{\rm d}r^2}{1-kr^2}+r^2({\rm d}\theta^2+\sin^2\theta
{\rm d}\varphi^2) \right],
\label{ds22}
\end{equation}
where in positive curvature (closed) universe ($k=+1$) $r\in\left<0,1\right>$, in
flat $(k=0)$ and negative curvature $(k=-1)$ open universes
$r\in\left<0,\infty\right>$, and $\theta\in (0,\pi)$, $\varphi\in(0,2\pi)$. We shall also employ
hyperspherical coordinates
\begin{equation}
ds^2={\rm d}t^2-a^2\left[  {\rm d}\chi^2+r^2({\rm d}\theta^2+\sin^2\theta
{\rm d}\varphi^2) \right],
\label{ds23}
\end{equation}
 with $r=\sin\chi$, $\chi$, $\sinh\chi$ for $k=1$, $0$, $-1$.

In a completely general gauge for
general perturbations  $h_{\mu\nu}$, the (momentum) constraint equation, 
$\delta G^0_k=\kappa \delta T^0_k$, turns out to be 
\begin{eqnarray}
\frac{1}{2}\nabla^2h_{0k}&+&kh_{0k}-\frac{1}{6}
a^2\nabla_k\nabla_j h_0^j+\frac{2}{3}a\nabla_k{\cal K} -\frac{1}{2}
a^2\dot{\cal T}_k
\nonumber
\\
&=&a^2\kappa\delta T^0_k,
\label{h0k}
\end{eqnarray}
where the dot denotes $\partial/\partial t$,
\begin{equation}
{\cal K}=a\left[  \frac{3}{2} H h_{00}-\frac{1}{2}
(h^j_j)^{\,\bf\dot{}}+\nabla_jh^j_0\right] 
\label{K}
\end{equation}
is the perturbed mean external curvature of $t={\rm constant}$ slices, 
$H=\dot a/a$
 is the Hubble constant,
\begin{equation}
{\cal T}_k=-\nabla_j\left(h^j_k-\frac{1}{3}\delta^j_kh^i_i\right).
\label{T}
\end{equation}
Notice that equation (\ref{h0k}) is independent of the choice of the
cosmological constant $\Lambda $ because we perturbed ``mixed"
components of $G^0_k$. Other perturbed Einstein's equations will not be
needed here. Since, however, we are interested primarily in perfect fluid
perturbations  we shall also consider the perturbed fluid equations  of motion, i.e. the
perturbed Bianchi identities  
\begin{eqnarray}
(\delta\rho)^{\,\bf\dot{}}+3H(\delta\rho+\delta
p)&+&
\\\nonumber
(\rho+p)\nabla_k(h_0^k+V^k)&+&(\rho+p)\left(\frac{3}{2}Hh_{00}-
\frac{1}{a}{\cal K}\right)=0,
\label{delta rho}
\end{eqnarray}
and 
\begin{eqnarray}
\frac{1}{a^3}\left[ 
a^3(\rho+p)(a^2f_{km}V^m-h_{0k})\right]^{\,\bf\dot{}}&+&\\\nonumber
\nabla_k\delta p+
\frac{1}{2}(\rho+p)\nabla_k h_{00}&=&0,
\label{Bianchi}
\end{eqnarray}
where $V^k=\frac{{\rm d}x^k}{{\rm d}t~}\simeq
\delta U^k$ and
$V_k=-a^2f_{kj}V^j$ is the fluid (small) velocity. The perturbed fluid
energy-momentum tensor components entering the constraint equations 
(\ref{h0k}) read
\begin{equation}
\delta T^0_k=(\rho+p)(h_{0k}+V_k)=(\rho+p)(h_{0k}-a^2f_{km}V^m).
\label{deltaT}
\end{equation}
There have been various choices of gauges used in the literature, in
particular the synchronous gauge $(h_{00}=h_{0k}=0)$. In order to
understand the effect of dragging of inertial frames, in particular its
`instantaneous' character, it is convenient to use gauges --- we call them
`Machian' --- in which the constraint equations, and still another
(combination of) the perturbed field equations  are explicitly the elliptic
equations.  In order to achieve this it is first useful to choose coordinates on
$t={\rm constant}$ slices such that the   {\it spatial} harmonic gauge
conditions are satisfied, i.e. ${\cal T}_k=0$, where ${\cal T}_k$ is given in
(\ref{T}) (in numerical relativity $\dot{\cal T}_k=0$ is frequently called the
`minimal distortion' shift vector gauge condition). Next, it is convenient
to choose the time slices so that, for example, the perturbation  of their external
curvature vanishes: ${\cal K}=0$, ${\cal K}$ given by (\ref{K}) (so called
`uniform Hubble expansion gauge'). Under these gauge conditions
(which determine the coordinates in a substantially more restrictive way
than e.g. the synchronous gauge) the constraint field equations  (\ref{h0k})
become the elliptic equations  for just the components $h_{0k}$, no other
$h_{\mu\nu}$ enter.

Until now we considered general perturbations  in the chosen gauge. Hereafter,
we assume the vectors $h_{0k}, V^k$ to be transverse,
\begin{equation}
\nabla^kh_{0k}=0,\qquad \nabla_kV^k=0,
\label{transverse}
\end{equation}
so that also $\nabla^k\delta T^0_k=0$. If (\ref{transverse}) is not satisfied,
we can apply $\nabla^k$ to equation  (\ref{h0k}), find the elliptic equation  for the
scalar $\nabla^kh_{0k}$, solve it and substitute back into (\ref{h0k}) where
the third term on the left hand side could be viewed as the source together
with $\delta T^0_k$. Since, however, the longitudinal parts do not
contribute to the dragging of inertial frames, we assume equations  (\ref{transverse}) to
be satisfied.

The constraint field equations  (\ref{h0k}) with our choice of gauge
${\cal K}={\cal T}_k=0$ [cf. equations  (\ref{K}) and (\ref{T})] thus become
\begin{equation}
\nabla^2h_{0k}+2kh_{0k}=2a^2\kappa\delta T^0_k=2a^2\kappa
(\rho+p)(h_{0k}-a^2f_{km}V^m),
\label{h0k2}
\end{equation}
where  for the perfect fluid $\delta T^0_k$ is substituted from equation 
(\ref{deltaT}). This is our basic equation  to be solved at a given time
$t={\rm constant}$, with either $\delta T^0_k$ or $V^k$ given. The Bianchi
identities (fluid equations  of motion) determine the time evolution of perturbations,
the scalar equation  (\ref{delta rho}) for $\delta\rho$, whereas the vector equation 
(\ref{Bianchi}) governs the evolution of the term
\begin{equation}
j_k\equiv  a^3(\rho+p)(a^2f_{km}V^m-h_{0k}) =-a^3\delta T^0_k.
\end{equation}
In the following we shall often express the background time dependent
term $a^2(\rho+p)$ by using equation (\ref{la}).

Consider first the flat universe ($k=0$). In Cartesian coordinates $x^k$
used by Schmid \cite{Sc}, the 3-metric $f_{kl}=\delta_{kl}$, and
(\ref{h0k2}) becomes
\begin{equation}
\nabla^2h_{0k}=2a^2\kappa\delta T^0_k=2a^2\kappa
(\rho+p)(h_{0k}+V_k), 
\label{h0k3}
\end{equation}
where $\nabla^2$ is the flat-space Laplacian. Substituting from equation 
(\ref{la}) with $k=0$ in the first term in the r.h.s. of equation  (\ref{h0k3}), we get 
\begin{equation}
\nabla^2h_{0k}=-4a^2\dot Hh_{0k}-2\kappa a^4(\rho+p)V_k.
\label{h0k4}
\end{equation} 

Now comparing our general form of the perturbed FRW metric with the
perturbed metric (5) in Schmid's work (and bewaring of the opposite
signature), we see that $h_{0k}=-a\beta_k$(Schmid). Considering
$(\rho+p)V_k$ (denoted by ${\vec J}_\epsilon$ in Schmid) as the source, the
equation  (\ref{h0k4}) written for Schmid's $\beta_k$ becomes
\begin{equation}
-\nabla^2\beta_k-4a^2\dot H\beta_k=-2\kappa a^3(\rho+p)V_k.
\end{equation}
This is Schmid's basic equation  (14), up to the factors $a^2$ and $a^3$
which in Schmid's equation  (14) are missing but this does not change
significantly Schmid's conclusions.

When $\delta T^0_k$ is given, the solution of equation  (\ref{h0k3})
is given as the Poisson integral over the source. If, however, the matter
current is given, equation  (\ref{h0k4}) can be written as
\begin{equation}
\nabla^2h_{0k}-\lambda^2(t)h_{0k}=-2\kappa a^4(\rho+p)V_k,
\label{h0k5}
\end{equation}
with ($k=0$)
\begin{equation}
\lambda^2=-4a^2\dot H.
\end{equation}
Usually (e.g. in the standard Friedmann models)  $\dot H<0$, so $\lambda$ is
real. The three equations   (\ref{h0k5}) are, as emphasized by Schmid, of the
Yukawa-type. The Green's functions are given by
\begin{equation}
G(x,x')= -\frac{1}{4\pi} \frac{e^{\mp\lambda|x-x'|}}{|x-x'|}
\label{G},
\end{equation}
the well-behaved solution of equation  (\ref{G}) is thus
\begin{equation}
h_{0k}=-\frac{1}{2\pi}\kappa a^4(\rho+p)\int
V_k(x')\frac{e^{-\lambda|x-x'|}}{|x-x'|}\,{\rm d}x'.
\end{equation}
Clearly {\it if\/} the perturbation  $V_k(x')$
 is located at $|x-x'|{\stackrel{>}{_\sim}}\lambda^{-1}=1/2a\sqrt{-\dot H}$,
i.e. beyond the `$\dot H ~$radius' $R_{\dot H}=2(-\dot H)^{-\frac{1}{2}}$ in
Schmid's terminology, the vector $h_{0k}$ which determines the dragging
of inertial frames  is exponentially suppressed around the origin. Although we thus
verified the interesting conclusion of Schmid, we do not resonate with his
view that `` because of the exponential cut-off... there is no need to impose
`appropriate boundary conditions of some kind'....". The Green's function in
(\ref{G}) with the `$+$' sign in the exponential is also the solution of equation 
(\ref{h0k5}) with a $\delta$-function source but one discards it by
demanding `reasonable' boundary conditions at infinity.

From the Machian viewpoint the closed universes are of course
preferable. There is, however, no vector Green's function available for equation 
(\ref{h0k2}) with either $\delta T^0_k$ or $V_k$ considered as a source.
In order to understand how Schmid's conclusions get modified in curved
universes and to generalize our previous work \cite {LKB95} which
analyzed perturbations  corresponding to rigid rotating spheres in the FRW
universes, we shall study all axisymmetric, odd-parity $l$-pole perturbations 
corresponding to differentially rotating `spheres'. 
We now derive the basic equations for such "toroidal" perturbations.
Their solutions, in particular for $l\geq 2$ and closed universes,
require special treatment. These solutions are analyzed in the following 
Paper~II.

In spherical coordinates [as in the FRW metrics (\ref{ds2}), (\ref{ds22})], the only
non-vanishing vector components are $h_{0\varphi}(t,r,\theta)$ and $V_\varphi
(t,r,\theta)$. [For  the general axisymmetric even-parity vector fields
$V_\varphi=0$, whereas  $V_r
(t,r,\theta)$ and $V_\theta
(t,r,\theta)$ are non-vanishing, the same being true for $h_{0r},
h_{0\theta}$].
There is now just one non-trivial constraint equation  in (\ref{h0k2}) to be
satisfied:
\begin{equation}
\nabla^2h_{0\varphi}+2kh_{0\varphi}=2a^2\kappa\delta T^0_\varphi,
\label{h0k6}
\end{equation}
in which $\nabla^2=f^{kl}\nabla_k\nabla_l$, with $f^{kl}$ being the inverse to
$f_{kl}$ given by FRW metric (\ref{ds22})
(recall -- see (\ref{ds2}) -- that $f_{kl}$ is positive definite, without factor
$a^2$). Calculating $\nabla^2h_{0\varphi}$ explicitly, we find equation  (\ref{h0k6})
to take the form
\begin{eqnarray}
&&\left[ (1-kr^2) \frac{\partial^2}{\partial r^2}-kr\frac{\partial}{\partial r} \right] h_{0\varphi} +
\\\nonumber
&+&\frac{1}{r^2} \sin \theta \frac{\partial}{\partial\theta}\left( \frac{1}{\sin\theta} 
\frac{\partial}{\partial\theta}\right) h_{0\varphi}+
4kh_{0\varphi}=2a^2\kappa\delta T^0_\varphi.
\label{constraint}
\end{eqnarray}
Before solving this constraint equation  it is interesting to notice what the
perturbed equations  of motion (Bianchi identities) say for  axisymmetric
odd-parity perturbations. Equation (\ref{delta rho}) in our gauge choice (with
${\cal K}=0$) and transverse character of $h_{0k}, V_k$
is a simple evolution equation  for $\delta\rho$. The vector equation  (\ref{Bianchi})
for indices $1,2 ~(x^1=r, x^2=\theta)$
 turns into the well known relativistic equilibrium conditions for perfect
fluids, $\nabla_k\delta p=-(\rho+p)\nabla_k (\frac{1}{2} h_{00})$ (see e.g. \cite{MTW}).
In the following the crucial role is played by equation   (\ref{Bianchi}) for
index $3 ~(x^3=\varphi)$. Since in axisymmetric case $\nabla_\varphi\delta p =0, 
\nabla_\varphi h_{00}=0$, it becomes   
\begin{equation}
\left[   a^3(\rho+p)\left(   a^2r^2\sin^2\theta\/ 
V^\varphi-h_{0\varphi}\right)\right]^{\,\bf\dot{}}=0,
\label{Bianchi0}
\end{equation}
or
\begin{equation}
\left[   a^3\delta T^0_\varphi\right]^{\,\bf\dot{}}=0.
\label{Bianchi2}
\end{equation} 
This is the conservation of angular momentum  of each element of each axially symmetrical
ring of fluid. The total angular momentum  in a spherical layer $\left<\chi_1,\chi_2\right>$
is given by
\begin{equation}
J(\chi_1,\chi_2)=-\int_{\chi_1}^{\chi_2}\!\!\!{\rm d}\chi\int_0^\pi
\!\!\!{\rm d}\theta\int_0^{2\pi}\!\!\!{\rm d}\varphi\sqrt{-\overline g}\,\delta T^0_\mu\eta^\mu,
\end{equation}
where $\eta^\mu=(0,0,0,1)$ is the rotational Killing vector, the
background metric determinant $\overline g=\overline
g^{(3)}=-a^6r^4\sin^2\theta, ~r=\sin\chi, \chi, \sinh\chi$ for
respectively $k=+1,0,-1$ as in equation (\ref{ds23}). Integrating  over $\varphi$
we have
\begin{eqnarray}
J(\chi_1,\chi_2)&=&-2\pi\int_{\chi_1}^{\chi_2}{\rm d}\chi\int_0^\pi
{\rm d}\theta ~a^3r^2\sin\theta\delta
T^0_\varphi
\nonumber\\
&=&2\pi \int_{\chi_1}^{\chi_2}{\rm d}\chi\int_0^\pi {\rm d}\theta~j(\theta,\chi,t),
\label{calJ} 
\end{eqnarray}
where $j(\theta,\chi,t)$ is the (coordinate) angular momentum  density. Hence,
the Bianchi identity (\ref{Bianchi2}) can be written as 
\begin{equation}
\left[  j(\theta,\chi)\right]^{\,\bf\dot{}}=0.
\label{Bianchi3}
\end{equation}
This is important for studying the time evolution of the $h_{0k}$ and
$V_k$ perturbations.

Defining the fluid angular velocity  
\begin{equation}
\Omega=V^\varphi=\frac{{\rm d}\varphi}{{\rm d}t},
\end{equation}
we get 
\begin{equation}
V_\varphi=-a^2f_{\varphi\varphi}V^\varphi=-a^2r^2\sin^2\theta~\Omega(t,r,\theta).
\end{equation}
Writing similarly
\begin{equation}
h_{0\varphi}=a^2r^2\sin^2\theta~\omega(t,r,\theta),
\end{equation}
the only non-vanishing component of $\delta T^0_k$ becomes 
\begin{equation}
\delta T^0_\varphi=(\rho+p) a^2r^2\sin^2\theta(\omega-\Omega).
\label{230}
\end{equation}
The angular momentum  density conservation law (\ref{Bianchi2}), resp.
(\ref{Bianchi3}), turns then into the simple evolution equation 
\begin{equation}
\left[   a^5(\rho+p)(\omega-\Omega)\right]^{\,\bf\dot{}}=0.
\label{Bianchi4}
\end{equation}

Let us now return back to the constraint equation  (\ref{constraint}). The
second term on its left hand side  suggests the decomposition into the
vector spherical harmonics. It should be emphasized
that, in contrast to standard practice in the cosmological  perturbation  theory where
perturbations  are decomposed into harmonics in all three spatial dimensions (see e.g. \cite{KS}), we
decompose in the usual coordinates $\theta,\varphi$ on spheres only, and
assume axial symmetry (spherical functions $Y_{lm}$ having $m=0$).
Thus, we write ($Y_{l0,\theta}\equiv\partial_\theta Y_{l0}$)
\begin{eqnarray}
h_{0\varphi}&=&a^2r^2\sum\limits_{l=1}^{\infty}\omega_l(t,r)\sin\theta\/ Y_
{l0,\theta},
\\V_\varphi&=&-a^2r^2\sum\limits_{l=1}^{\infty}\Omega_l(t,r)\sin\theta\/ Y_{l0,\theta},
\end{eqnarray} 
and
\begin{eqnarray}
\delta
T^0_\varphi&=&a^2(\rho+p)r^2\sum\limits_{l=1}^{\infty}
(\omega_l-\Omega_l)\sin\theta~ Y_{l0,\theta}
\nonumber\\
&=&
\sum\limits_{l=1}^{\infty}[\delta
T^0_\varphi(t,r)]_l\sin\theta\/ Y_{l0,\theta}.
\label{234}
\end{eqnarray}
Substituting these expansions into equation  (\ref{constraint}) and using the
orthogonality of functions $\sin\theta ~Y_{l0,\theta}$  for different $l$'s, we
obtain the `radial' equation  for each $l$:
\begin{eqnarray}
&&\left[ (1-kr^2) \frac{\partial^2}{\partial r^2}-kr\frac{\partial}{\partial r} \right] (r^2\omega_l)
-l(l+1)\omega_l+4kr^2\omega_l
\nonumber\\
&&=2a^2\kappa
(\rho+p)r^2(\omega_l-\Omega_l)=\lambda^2r^2(\omega_l-\Omega_l),
\label{constraint2}
\end{eqnarray}
where we used equation (\ref{la}).
It is easy to convert the last equation  into the form
\begin{eqnarray}
-\sqrt{1-kr^2}\frac{1}{r^2}\frac{\partial}{\partial r}\left[ 
\sqrt{1-kr^2}\frac{\partial}{\partial r}(r^2\omega_l)
\right]+
\nonumber\\
\frac{l(l+1)}{r^2}\omega_l-4k\omega_l=\lambda^2(\Omega_l-\omega_l).
\label{oml}
\end{eqnarray} 
For $l=1$ (and the background pressure $p=0$) this equation  coincides
exactly with equation  (4.32) in LKB. In the language of the present paper, in
LKB we analyzed dipole $(l=1)$ axisymmetric odd-parity perturbations. With
$l=1$, $Y_{10,\theta}=-\sqrt{3/4\pi}\sin \theta$, so that putting
$\omega=-\sqrt{3/4\pi }~\omega_{l=1}$, $\Omega=-\sqrt{3/4\pi }~\Omega_{l=1}$,
we recover 
\begin{eqnarray}
h_{0\varphi}&=&a^2r^2\sin^2\theta~\omega(t,r),\quad V_\varphi=-a^2r^2\sin^2\theta~
\Omega(t,r),\nonumber\\
\delta T^0_\varphi&=&a^2(\rho+p)r^2\sin^2\theta(\omega-\Omega),
\label{237}
\end{eqnarray}
which corresponds to the {\it rigidly} rotating spheres in the FRW
universes considered in Section 4.4 in LKB, and, for $\Omega(t,r)$ given,
analyzed in detail in Section 4 in the following.

Consider first the case $k=0$. Equation (\ref{oml}) can be written with the
angular momentum  density $(\delta T^0_\varphi)_l$ as a source,
\begin{equation}
\frac{1}{r^4}\frac{\partial}{\partial r}\!\!\left(\!r^4 \frac{\partial\omega_l}{\partial
r}\!\right)-\frac{l(l+1)-2}{r^2}\omega_l=\lambda^2(\omega_l-\Omega_l)
=\frac{2\kappa}{r^2}(\delta
T^0_\varphi)_l.
\label{oml8}
\end{equation}
If the fluid angular velocity  is taken as a source, the equation  reads 
\begin{equation}
\frac{1}{r^4}\frac{\partial}{\partial r}\left(  r^4 \frac{\partial\omega_l}{\partial
r}
\right)-\left[\lambda^2+\frac{l(l+1)-2}{r^2}\right]
\omega_l=-\lambda^2\Omega_l,
\label{oml2}
\end{equation}
where $\lambda^2=-4a^2\dot H=2\kappa a^2(\rho+p)$ by using equation  (\ref{la})
with $k=0$.

In the case of spatially curved $(k\ne 0)$ backgrounds it is advantageous
to write $r^2=k(1-\mu^2)$, i.e. $\mu=\sqrt{1-kr^2}$ to
obtain
%
%
\begin{eqnarray}
\frac{1}{[k(1-\mu^2)]^{3/2}}\frac{\partial}{\partial\mu}\Big\{ 
[k(1-\mu^2)]^{5/2}\frac{\partial\omega_l}{\partial\mu}\Big\}
-\frac{l(l+1)-2}{k(1-\mu^2)}\omega_l&\!\!&\!\!\nonumber\\
=\frac{2\kappa}{k(1-\mu^2)}(\delta
T^0_\varphi)_l.\hphantom{AAAAA}&\!\!&\!\!\label{oml3}
\end{eqnarray}
The substitution
\begin{equation}
\omega_l=[k(1-\mu^2)]^{-3/4}\overline \omega_l
\label{241}
\end{equation}
turns equation  (\ref{oml3}) into the Legendre equation  for $\overline\omega_l$ with
$(\delta T^0_\varphi)_l$
 as the source:
\begin{eqnarray}
&&\frac{\partial}{\partial\mu}\left[
k(1-\mu^2)\frac{\partial\overline\omega_l}{\partial\mu}\right]+\left[ k \frac{3}{2}
(\frac{3}{2}+1)-
\frac{(l+\frac{1}{2})^2}{k(1-\mu^2)}
\right] \overline\omega_l
\nonumber\\
&&\hphantom{AAAAAAAAAAAA}
=\frac{2\kappa}{[k(1-\mu^2)]^{1/4}}(\delta T^0_\varphi)_l.
\label{242}
\end{eqnarray}
Finally, considering the fluid angular velocity  as the source, we can write the last
equation  again as the Legendre equation  with a more complicated degree:
\begin{eqnarray}
&&\frac{\partial}{\partial\mu}\left[ 
k(1-\mu^2)\frac{\partial\overline\omega_l}{\partial\mu}\right]+\left[ k\nu(\nu+1)-
\frac{(l+\frac{1}{2})^2}{k(1-\mu^2)}
\right]\overline\omega_l
\nonumber\\
&&\hphantom{AAAA}
=-K_l\equiv-\lambda^2\Omega_l[k(1-\mu^2)]^{3/4},
\label{oml4}
\end{eqnarray}
where
\begin{equation}
\left(\nu+\frac{1}{2}\right)^2=4-2k\kappa a^2(\rho+p)=4-k\lambda^2=4ka^2\dot H.
\label{nu}
\end{equation}
The degree $\nu$ of the Legendre equation  does not depend on $l$. For
$l=1$, equation  (\ref{oml4}) goes over into equation  4.35  in LKB \cite{foot2}.

\sect{Solutions for $\omega$ with given angular momentum  distribution }

We shall start by making more explicit the solutions obtained in LKB
which are the $l=1$ odd-parity vector solutions of the general problem.
In such modes each sphere rotates with no shear but it expands (or
contracts) with the background  and as it does so its angular velocity  changes (see Section 5).

\medskip
\noindent \underline{$k=0$}

\nobreak
The equation  to be solved is (\ref{oml8}) with $l=1$, this is 4.33 LKB
\begin{equation}
\frac{1}{r^4}\frac{\partial}{\partial r}\left(  r^4 \frac{\partial\omega}{\partial r}
\right)=-\lambda^2(\Omega-\omega)=\frac{2\kappa}{r^2}\delta T^0_\varphi,
\label{om}
\end{equation}
multiplying up by $r^4$
this takes the form
\begin{equation}
\frac{\partial}{\partial r}\left(  r^4
\frac{\partial\omega}{\partial r}\right)=-\frac{6}{a^3}\frac{{\rm d}J(<r)}{{\rm d}r},
\end{equation}
so
\begin{equation}
\frac{\partial\omega}{\partial r}=-\frac{6J}{a^3r^4},
\end{equation}
the constant of integration is zero since $J(<r)$ is zero at $r=0$
 where $\partial\omega/\partial r$ must vanish. Integrating again and insisting that
$\omega\rightarrow 0$ at $\infty$
we find
\begin{eqnarray}
\omega&=&a^{-3}\!\int_r^\infty\!\frac{6J}{r{'}^4}\,{\rm d}r'=2a^{-3}\left[ 
\frac{J(<r)}{r^3}+\!\int_r^\infty\!\!\frac{{\rm d}J}{{\rm d}r'}r{'}^{-3}\,{\rm d}r'
\right]\nonumber
\\&=&\frac{2}{r^3} \int_0^r\!\!\!\int_0^\pi 2\pi r{'}^2\sin\theta(-\delta
T^0_\varphi)\,{\rm d}\theta\,{\rm d}r'
\nonumber\\
&&
+ 2\int_r^\infty\!\!\!\int_0^\pi 2\pi r{'}^{-1}\sin\theta(-\delta
T^0_\varphi)\,{\rm d}\theta\,{\rm d}r',
\label{0omJ}
\end{eqnarray}
where we have used (\ref{calJ}) to define $J(<r)$ in terms of $\delta T^0_\varphi$.

\medskip
\noindent \underline{$k=1$}

\nobreak
The equation  to be solved is (\ref{oml3}) with $l=1$ which is 4.34
LKB \cite{foot3}
\begin{eqnarray}
\frac{\partial}{\partial\mu}\Big\{ 
(1-\mu^2)^{5/2}\frac{\partial\omega}{\partial\mu}\Big\}&=&2\kappa(1-\mu^2)^\frac{1}{2}\delta
T^0_\varphi= \frac{6}{a^3}\frac{{\rm d}J}{{\rm d}\mu}
\nonumber\\
\noalign{\noindent so that}
(1-\mu^2)^\frac{1}{2}\frac{\partial\omega}{\partial\mu}&=&\frac{6J}{a^3(1-\mu^2)^2}.
\end{eqnarray}
As before there is no integration constant for the same reason. We now
write $\mu=\cos \chi$, then $\chi$ is the normal cosmic radial angle and
\begin{equation}
\frac{\partial\omega}{\partial\chi}=-\frac{6J}{a^3\sin^4\chi}.
\label{36}
\end{equation}
Now
\begin{equation}
\int^\chi\frac{{\rm d}\chi}{\sin^4\chi}=-\frac{1}{3}  
\left(\cot^3\chi+3\cot\chi\right)=-\frac{1}{3} W(\chi).
\label{37}
\end{equation}
Hence
\begin{equation}
\omega=2a^{-3}\left[ WJ(<\chi)+\int_\chi^\pi
W\frac{{\rm d}J}{{\rm d}\chi'}\,{\rm d}\chi'\right] +\omega_0,
\label{1omJ}
\end{equation}
where
\begin{equation}
J=\int_0^\chi\!\!\int_0^\pi 2\pi a^3[r(\chi')]^2\sin^2\theta(-\delta
T^0_\varphi)\,{\rm d}\theta\,{\rm d}\chi'.
\label{J}
\end{equation}

Just as in the last case $W$ diverges at $\chi=0$
like $\chi^{-3}$, however, the angular momentum  of spheres near the origin is sufficiently
small to make the $WJ$ tend to a constant as $\chi$
tends to zero. It is shown in LKB that the condition of convergence of the
second integral at  $\chi=\pi$
is that the total angular momentum  of the universe is zero. If that condition is fulfilled
and $\Omega-\omega$ is regular near $\chi=\pi$ then the integral converges.
If the total angular momentum  is not zero then the integral for $\omega$
diverges at $\chi=\pi$. Thus for $\omega$
to be finite at $\chi=\pi$ the total angular momentum  must be zero in the closed
universe. There is no way of fixing $\omega_0$
because there is no standard of zero rotation, as there is for the infinite
universes. Indeed, according to Mach a description of the world in
rotating axes is just as good in principle as a description in non-rotating
ones. Note that the source $\Omega-\omega$ does not change when the axes
are rotating since $\Omega$ and $\omega$ acquire the same constant $\omega_0$.
An absolute rotation can arise only from spatial boundary conditions
which do not occur for closed universes.

\medskip
\noindent\underline{$k=-1$}

\nobreak
The equation  to be solved is (\ref{oml3}) with $l=1$. Multiplying through by
$(\mu^2-1)^{3/2}$
 we  obtain
\begin{equation}
\frac{\partial}{\partial\mu}\Big\{ 
(\mu^2-1)^{5/2}\frac{\partial\omega}{\partial\mu}\Big\}=-2\kappa \sqrt{\mu^2-1}\delta
T^0_\varphi=-\frac{6}{a^3}\frac{{\rm d}J}{{\rm d}\mu},
\end{equation}
so on integration and division
\begin{equation}
(\mu^2-1)^{1/2}\frac{\partial\omega}{\partial\mu}=-\frac{6}{a^3}(\mu^2-1)J.
\end{equation}
Writing $\mu=\cosh\chi$ to introduce the natural radial variable of
hyperbolic space, this becomes
\begin{equation}
\frac{\partial\omega}{\partial\chi}=-\frac{6}{a^3}\frac{J}{\sinh^4\chi}.
\end{equation}
 Integrating again and insisting that $\omega\rightarrow 0$
at infinity we use the integral
\begin{equation}
\int^\chi\frac{{\rm d}\chi'}{\sinh^4\chi'}=-\frac{1}{3}\left(  
\coth^3\chi-3\coth\chi+2\right)\equiv-\frac{1}{3}\overline W(\chi),
\label{313}   
\end{equation}
and on integrating by parts we obtain 
\begin{equation}
\omega=2a^{-3}\left[ 
\overline W J(<\chi)+\int_{\chi}^{\infty}\overline W(\chi')\frac{dJ}{{\rm d}\chi'}{\rm d} \chi'\right],
\label{-1omJ}
\end{equation}
where $J$ is the same as in (\ref{J}) with $r=\sinh\chi$. We have chosen
the above definition of $\overline W$ so that $\overline W\rightarrow 0$
at infinity; so no constant of integration is needed to incorporate the
boundary condition that $\omega\rightarrow 0$.

\sect{Solutions for $\omega$ with given $\Omega$ }

The method of solution was outlined in LKB but here we work through all
the details starting with the simplest case. 

\medskip
\noindent\underline {$k=0$}

The relevant equation  to be solved is (\ref{oml2}) with $l=1$, equation  4.33 in
LKB,  rewritten as
\begin{equation}
\frac{1}{r^4}\frac{\partial}{\partial r}\left(  r^4 \frac{\partial\omega}{\partial r}
\right)-\lambda^2\omega=-\lambda^2\Omega.
\label{om2}
\end{equation}
Here $\lambda^2=2a^2\kappa(\rho+p)>0$. $\lambda^{-1}a$ has the units of a length
and we shall call it, following Schmid \cite{Sc}, the distance to the horizon. 
In dimensionless comoving
coordinates this corresponds to $r=\lambda^{-1}$. We write $z=\lambda r$ and
$\partial\omega/\partial z=\omega'$. Then equation  (\ref{om2})  reduces to
\begin{equation}
\omega''+\frac{4}{z}\omega-\omega=-\Omega.
\label{42}
\end{equation}
The corresponding homogeneous equation  is Bessel's equation  for $z^{-3/2}J_{3/2}(iz)$,
 which has real solutions  
$\omega=\overline
{\cal I}$ and
$\omega=\overline
{\cal K}$, where
$\overline{\cal I}=z^{-3/2}I_{3/2}(z)$ and $\overline{\cal K}=z^{-3/2}K_{3/2}(z)$.
For small $z$,
$\overline{\cal I}\rightarrow\frac{1}{3}\sqrt{2/\pi}(1+z^2/10)$; $\overline{\cal K}\rightarrow\sqrt{\pi/2}~
z^{-3}$. For large $z$, $\overline{\cal I}\rightarrow(1/\sqrt{2\pi})z^{-2}e^z$; 
$\overline{\cal K}\rightarrow\sqrt{\pi/2}~z^{-2}e^{-z}$.

We use the method of variation of parameters to solve the
inhomogeneous equation  with boundary conditions that $\omega$ tends to zero
at infinity and to a constant at the origin. We thus obtain
\begin{eqnarray}
\omega(z)=&&\overline{\cal K}(z)\int_0^z (z')^4\overline{\cal I}(z')\Omega(z')\,{\rm d}z'
\nonumber\\
&+&\overline{\cal I}(z)
\int_z^\infty
(z')^4\overline{\cal K}(z')\Omega(z')\,{\rm d}z'.
\label{0om}
\end{eqnarray} 
For the solutions near the origin with sources that are not so close, we
may neglect the first term and then for small $z$,
\begin{equation}
\omega(z)=\frac{1}{3}\sqrt{\frac{2}{\pi}}\left(1+\frac{z^2}{10}\right)\int_z^
\infty(z')^4\overline{\cal K}(z')\Omega(z')\,{\rm d}z'.
\label{44}
\end{equation}
When the source $\Omega$ is beyond the horizon $z=1$, i.e. $z'\gg1$,
\begin{equation}
\omega(z)=\frac{1}{3}\left(1+\frac{z^2}{10}\right)\int_z^
\infty(z')^2e^{-z'}\Omega(z')\,{\rm d}z';
\end{equation}
so for a source localized in $r_0(1\pm\Delta)$ with $\Delta\ll1/\lambda$,
\begin{equation}
\omega(z)=\frac{1}{3}\left(1+\frac{\lambda^2r^2}{10}\right)(\lambda r_0)^3e^{-\lambda r_0}\overline\Omega 2\Delta,
\label{46}
\end{equation}
which clearly shows the exponential decline of influence remarked on by
Schmid \cite{Sc}. When $\Omega$
is concentrated near $z_0$, in $z_0\pm\lambda\Delta$, then with $z_0\gg 1$ and $\Omega=\overline\Omega$ we get
\begin{equation}
\omega(z_0)=\frac{1}{2}\int_0^\infty\!\left(\frac{z'}{z_0}\right)^2e^{-|z'-z_0|}\Omega(z')\,{\rm d}z'\simeq
\lambda\Delta\overline\Omega.
\label{47}
\end{equation}
Thus {\it at} the source the inertial frame  rotates at $\lambda\Delta\overline\Omega$ and
$\overline\Omega-\omega=(1-\lambda\Delta)\overline\Omega$. 

We now turn to the solutions for a closed universe.

\medskip
\noindent\underline{$k=1$}

The relevant equation  is Legendre's equation  for $\overline\omega=(1-\mu^2)^{3/4}\omega
$ with an inhomogeneous term written below. This is LBK equation  4.35 and
the same as the equation  (\ref{oml4}) of Section 2 of this paper specialized for
$l=1$:
\begin{eqnarray}
\frac{\partial}{\partial\mu}\Big\{ 
(1-\mu^2)\frac{\partial\overline\omega}{\partial\mu}\Big\}&+&\Big\{\nu(\nu+1)-
\frac{(\frac{3}{2})^2}{(1-\mu^2)}
\Big\}\overline\omega=-K,
\nonumber\\
&&K\equiv\lambda^2\Omega(1-\mu^2)^{3/4},
\label{48}
\end{eqnarray}
where $(\nu+\frac{1}{2})^2=4-\lambda^2$ as in (\ref{nu}). Since $k=+1$ the space is
hyperspherical and the convention is to write $\mu=\cos\chi$
so that $\chi$ becomes the radial variable. The solutions of the
homogeneous equation  are the Legendre functions $P^{3/2}_\nu(\mu)$ and
$Q^{3/2}_\nu(\mu)$ and a recurrence relation that generates
$P^{\mu+1}_\nu$ from $P^\mu_\nu$ and $P^\mu_{\nu-1}$. 
(Here the order $\mu$ of the Legendre function has nothing to do
with the variable $\mu=\sqrt{1-kr^2}$.)
Thus
\begin{eqnarray}
P^{1/2}_\nu(\cos\chi)&=&\hphantom{-}\Big(\frac{\pi}{2}\Big)^{-\frac{1}{2}}(\sin\chi)^{-\frac{1}{2}}
\cos\Big[\Big(\nu+\frac{1}{2}\Big)\chi\Big],\nonumber\\
Q^{1/2}_\nu(\cos\chi)&=&-\Big(\frac{\pi}{2}\Big)^\frac{1}{2}(\sin\chi)^{-\frac{1}{2}}
\sin\Big[\Big(\nu+\frac{1}{2}\Big)\chi\Big].
\label{49}
\end{eqnarray}
To keep $P^{3/2}_\nu(\cos\chi)$ and
$Q^{3/2}_\nu(\cos\chi)$ real, we use $(1-\mu^2)^{-\frac{1}{2}}=(\sin\chi)
^{-1}$
 in place of $(\mu^2-1)^{-\frac{1}{2}}$ in the recurrence relation 8.5.1 of 
Abramowitz and Stegun \cite{AS} (this merely multiplies the results by $-i$).
 \begin{equation}
P^{3/2}_\nu(\cos\chi)=\frac{1}{\sin\chi}\!\left[\Big(\nu-\frac{1}{2}\Big) P^\frac{1}{2}_\nu\cos\chi
+\Big(\nu+\frac{1}{2}\Big)P^\frac{1}{2}_{\nu-1}\right],
\label{410}
\end{equation}
the same relation holds for the $Q^{3/2}_\nu$. It turns out to be
convenient to write $n=\nu+\frac{1}{2}$. We note that (\ref{nu}) and (\ref{48}) involve this
quantity and that $n$ can be real but is often imaginary. Thus
\begin{eqnarray}
&&P^{3/2}_{n-\frac{1}{2}}(\cos\chi)=\\\nonumber
&&-\left(\frac{\pi}{2}\right)^{-\frac{1}{2}}\frac{1}{\sin^{3/2}\chi}
\left[   
\cos\chi \cos(n\chi)-n\sin\chi\sin(n\chi)\right],
\end{eqnarray}
similarly writing $n$ when it is real but $n=iN$ when it is imaginary:
\begin{eqnarray}
&&Q^{3/2}_{n-\frac{1}{2}}(\cos\chi)=
\\
&&\hphantom{i}\Big(\frac{\pi}{2}\Big)^{\frac{1}{2}}\frac{1}
{\sin^{3/2}\chi}
\left[   
\cos\chi \sin(n\chi)-n\sin\chi\cos(n\chi)\right],
\nonumber\\
&&Q^{3/2}_{iN-\frac{1}{2}}(\cos\chi)=
\nonumber\\\nonumber
&&i\Big(\frac{\pi}{2}\Big)^{\frac{1}{2}}\frac{1}{\sin^{3/2}\chi}
\left[   
\cos\chi \sinh(N\chi)-N\sin\chi\cosh(N\chi)\right].
\label{412}
\end{eqnarray}
We shall be concerned to have functions which, after multiplication by
another $(\sin\chi)^{-3/2}$, are nevertheless still finite at the origin
$\chi=0$. A little expansion around $\chi=0$ shows that the
$P$ function diverges but $Q$ function satisfies this stringent test. Our
next job is to find a solution that satisfies this stringent convergence not
at $\chi=0$ but at the `other' $r=0$ at $\chi=\pi$. Since that is an
alternative origin it is clear that $Q^{3/2}_{n-\frac{1}{2}}[\cos(\pi-\chi)]$
passes that test. A little work shows that it is indeed the linear
combination
$(2/\pi)\sin(n\pi)P^{3/2}_{n-\frac{1}{2}}(\chi)-\cos(n\pi)
Q^{3/2}_{n-\frac{1}{2}}(\chi)$. Finally we notice that $n=0$, which is needed
in some of our solutions, gives $Q_{-1/2}^{3/2}\equiv 0$. This is not a
solution at all! However $\lim\limits_{n\rightarrow 0}[(1/n)Q_{n-1/2}^{3/2}]$
gives the finite limit
\begin{equation}
\Big(\frac{\pi}{2}\Big)^\frac{1}{2}\frac{1}{\sin^{3/2}\chi}\left[   \chi\cos\chi
-\sin\chi \right].
\end{equation}
We shall therefore use the functions
\begin{eqnarray}
q_n&=&\Big(\frac{\pi}{2}\Big)^{\frac{1}{2}}\frac{1}{\sin^{3/2}\chi}S_n(\chi)\;,
\\\nonumber
p_n&=&\Big(\frac{\pi}{2}\Big)^{\frac{1}{2}}\frac{1}{\sin^{3/2}\chi}S_n(\pi-\chi)
\label{414}
\end{eqnarray}
as our independent solutions of the Legendre equation. These functions have
the added advantage that they remain real when $n=iN$:
\begin{eqnarray}
S_n(\chi)&=&-\cos\chi 
\frac{\sin(n\chi)}{n}+\sin\chi\cos(n\chi),\\\nonumber
S_{iN}(\chi)&=&-\cos\chi 
\frac{\sinh(N\chi)}{N}+\sin\chi\cosh(N\chi).
\end{eqnarray}
The Wronskian may be shown to be
\begin{equation}
p_n\frac{{\rm d}q_n}{{\rm d}\mu}-q_n\frac{{\rm d}p_n}{{\rm d}\mu}=\frac{\pi}{2}\frac{\sin(n\pi)}{n}
\frac{n^2-1}{1-\mu^2}=\frac{{\cal W}}{1-\mu^2}.
\label{416}
\end{equation}
Having formed solutions $p$
and $q$ each of which satisfy {\it one} of the boundary conditions we
look for solutions of the inhomogeneous equation  of the form
\begin{equation}
\overline\omega=A(\mu)p+B(\mu)q.
\end{equation}
We choose $A'p+B'q=0$, and then the equation  demands that
\begin{equation}
(1-\mu^2)[A'p'+B'q']=-K,
\end{equation}
where a dash denotes $\partial/\partial\mu$. Solving for $A'$
and $B'$ we have, using the Wronskian ${\cal W}/(1-\mu^2)$ defined earlier,
$A'=Kq/{\cal W}$. Now $p$ does not satisfy the boundary conditions at
$\chi=0$, so $A$  must be zero there; hence
\begin{equation}
A=-\int_\mu^1\frac{Kq}{{\cal W}}\,{\rm d}\mu=-\int_0^\chi\frac{Kq}{{\cal W}}\sin\chi~
\,{\rm d}\chi.
\end{equation}
Similarly $B'=-Kp/{\cal W}$ and to satisfy the boundary conditions at $\mu=-1,
\chi=\pi$, 
\begin{equation}
B=-\int_{-1}^\mu\frac{Kp}{{\cal W}}\,{\rm d}\mu=-\int_\chi^\pi\frac{Kp}{{\cal W}}
\sin\chi\,{\rm d}\chi.
\end{equation}
Thus the solution by variation of the parameters is
\begin{equation}
\overline\omega=-\left[  p(\chi)\int_0^\chi\!\!\frac{Kq}{{\cal W}}\sin\chi'\,{\rm d}\chi'+q(\chi)
\int_\chi^\pi\!\!\frac{Kp}{{\cal W}}
\sin\chi'\,{\rm d}\chi'
 \right],
\end{equation}
which gives our solution for $\omega(\chi)=(\sin\chi)^{-3/2}\overline\omega$:
\begin{eqnarray}
\omega(\chi)=&-&\frac{\pi/2}{{\cal W}\sin^3\chi}\left[
S_n(\pi-\chi)\!\int_0^\chi\!\!\!\lambda^2\Omega S_n(\chi')\sin\chi'\,{\rm d}\chi'
\right.
\nonumber\\
&+&
\left.
S_n(\chi)\int_\chi^\pi\!\!\lambda^2\Omega S_n(\pi-\chi')\sin\chi'\,{\rm d}\chi'
\right].
\label{1om}
\end{eqnarray}
For $\chi$ small,
\begin{eqnarray}
S_n&\rightarrow&
\frac{(1-n^2)}{3}\chi^3\left[1-\frac{(1+n^2)\chi^2}{10}\right],~~{\rm i.e.}
\nonumber\\
\frac{1}{\sin^3\chi}S_n&\rightarrow& \frac{(1-n^2)}{3}\left[
1+\frac{(4-n^2)}{10}\chi^2\right],
\end{eqnarray}
and for $n=iN$,
\begin{equation}
\frac{1}{\sin^3\chi}S_{iN}\rightarrow
\frac{(1+N^2)}{3}\left[1+\frac{(4+N^2)}{10}\chi^2\right].
\end{equation}
We note that with $k=+1~,~4+N^2=\lambda^2$, and
\begin{equation}
{\cal W}=\frac{\pi}{2}(n^2-1)\frac{\sin(n\pi)}{n}=-\frac{\pi}{2}(1+N^2)\frac{\sinh(N\pi)}{N}.
\end{equation}
For $N$ large and $\chi$ small
\begin{equation}
\frac{S_{iN}}{{\cal W}\sin^3\chi}\rightarrow
-\frac{4}{3\pi}Ne^{-N\pi}\left( 1+\frac{\lambda^2\chi^2}{10}\right).
\end{equation}
For  $N$ large and $\chi$ not small nor near $\pi$,
\begin{equation}
S_{iN}(\chi)= \frac{1}{2} \sin\chi ~e^{N\chi},\qquad S_{iN}(\pi-\chi)= \frac{1}{2}
\sin\chi ~e^{N(\pi-\chi)}.
\end{equation}
Hence our solution near the origin is
\begin{equation}
\omega(\chi)=\frac{1}{3}\left( 1+\frac{\lambda^2\chi^2}{10}\right) N  \int_\chi^\pi
\lambda^2\Omega(\chi')\sin^2\chi' e^{-N\chi'}\,{\rm d}\chi',
\label{428}
\end{equation}
and near the perturbation 
\begin{equation}
\omega(\chi_0)=\frac{1}{2}\int_0^\pi
\frac{\lambda^2 N}{N^2+1}\left(\frac{\sin\chi'}{\sin\chi_0}\right)^2
e^{-N|\chi'-\chi_0|}\Omega(\chi')\,{\rm d}\chi',
\label{429}
\end{equation}
where at the last line we consider a perturbation  with a mean $\Omega$ of
$\overline\Omega$ in $r_0\pm\Delta$ with $N\Delta\ll1$.

\medskip
\noindent\underline{$k=-1$}

The equation  to be solved is (\ref{oml4}) with $k=-1$ and $l=1$. Now we write
$\mu=\cosh\chi~,~(\nu+\frac{1}{2})^2=\lambda^2+4$. Space is now hyperbolic and
$\mu$ runs from 1 to $\infty$. The relevant solutions of the
homogeneous equation  are
\begin{eqnarray}
p&=&-\left( P^{3/2}_\nu   +\frac
{2}{\pi}iQ^{3/2}_\nu\right)=\frac{1}{2}\Big(\frac{\pi}{2}\Big)^{-\frac{1}{2}}
\frac{1}{\sinh^{3/2}\chi}S_e(\chi),\nonumber\\
q&=& iQ^{3/2}_\nu=\frac{1}{2}\left(\frac{\pi}{2}\right)^{\frac{1}{2}}
\frac{1}{\sinh^{3/2}\chi}E(\chi),
\end{eqnarray}
where $n=(\nu+\frac{1}{2})$,
\begin{eqnarray}
E(\chi)&=&-(n-1)e^{-(n+1)\chi}+(n+1)e^{-(n-1)\chi},
\nonumber\\
S_e(\chi)&=&\frac{1}{2}\left[  E(\chi)-E(-\chi) \right].
\label{431}
\end{eqnarray}
The Wronskian  
\begin{equation}
p\frac{{\rm d}q}{{\rm d}\mu}-q\frac{{\rm d}p}{{\rm d}\mu}=-\frac{(n^2-1)n}{\mu^2-1}.
\end{equation}
The solution by variation of parameters is
\begin{equation}
\overline\omega=-\frac{1}{n(n^2-1)}\left[ 
p\int_\mu^\infty qK\,{\rm d}\mu+q\int^\mu_1pK\,{\rm d}\mu\right],
\end{equation}
hence, changing the integrations from $\mu$ to $\chi$
and $\overline \omega$  to $\omega$, we have
\begin{eqnarray}
\omega=&&\frac{(\sinh\chi)^{-3}}{4(n^2-1)n}\left[ E(\chi)\int_0^
\chi\lambda^2\Omega(\chi') S_e(\chi')\sinh\chi'\,{\rm d}\chi'
\right.
\nonumber\\
&+& \left. S_e(\chi)\int_\chi^\infty\lambda^2\Omega(\chi') E(\chi')\sinh\chi'\,{\rm d}\chi' 
\right].
\label{434}
\end{eqnarray}
For small $\chi$
\begin{eqnarray}
E(\chi)&=&2-(n^2-1)\chi^2\times
\\\nonumber
&&\left[   1-
\frac{2n}{3}\chi+\frac{3n^2+1}{12}\chi^2-\frac{n(n^2+1)}{15}
\chi^3+...\right],
\end{eqnarray}
so
\begin{eqnarray}
S_e(\chi)&=&\frac{2n}{3}(n^2-1)\chi^3\left[   1-
\frac{(n^2+1)}{15}\chi^2\right] .
\end{eqnarray}
At large $\chi$
\begin{eqnarray}
E(\chi)&=&(n+1)e^{-(n-1)\chi}=2(n+1)e^{-n\chi}\sinh\chi,\nonumber\\
S_e(\chi)&=& \frac{1}{2} (n-1)e^{(n+1)\chi}=(n-1)e^{n\chi}\sinh\chi.
\end{eqnarray}
Near the origin
\begin{eqnarray}
\omega&=&\frac{1}{3}\left[1-\frac{(4-n^2)\chi^2}{10}\right]\times
\\\nonumber
&&\int_\chi^\infty(n^2-4)(n+1)\sinh^2\chi'e^{-n\chi'}\Omega (\chi')\,{\rm d}\chi'.
\label{437}
\end{eqnarray}
At the perturbation 
\begin{equation}
\omega(\chi_0)=\frac{1}{2}\frac{n^2-4}{n}\int_0^\infty
\left(\frac{\sinh \chi'}{\sinh \chi_0}\right)^2
e^{-n|\chi'-\chi_0|}\Omega(\chi')\,{\rm d}\chi'.
\label{438}
\end{equation}

\sect{The time evolution of the dragging}

The evolution of $\omega$
and $\Omega$ as functions of cosmic time is governed by the equations of
motion (contracted Bianchi identities) (\ref{Bianchi}). For axisymmetric,
odd-parity perturbations  these become the angular momentum  density conservation law, as
discussed in equations  (\ref{Bianchi0})--(\ref{Bianchi3}) in Section 2. In terms
of $\omega(t,r,\theta)$ and $\Omega(t,r,\theta)$ the conservation law simply becomes
(\ref{Bianchi4}), i.e.
\begin{equation}
\left[   a^5(\rho+p)(\omega-\Omega) \right]^{\,\bf\dot{}}=0
\end{equation}
or, in terms of the angular momentum  density, we get
\begin{equation}
\Omega-\omega=\frac{1}{a^5(\rho+p)}\cdot\frac{j(\chi,\theta)}{r^4\sin^3\theta}.
\label{Om-om}
\end{equation}
In this formula the first factor singles out the time dependence of
$\Omega-\omega$. Notice that we have already obtained $\omega(t,r,\theta)$ as a
function of the angular momentum  within $\chi$, $J(<\chi)$, in all three cases
$k=+1,0,-1$ [see equations  (\ref{0omJ}), (\ref{1omJ}), (\ref{-1omJ})]. We found
$\omega$ to depend on the time as $1/a^3(t)$. Equation (\ref{Om-om}) then can
be regarded as a solution $\Omega(t,r,\theta)$ implied by the equations  of motion.

On the other hand, for $\Omega-\omega$ given at some time $t=t_0$ as a
function of $\chi,\theta$, equation  (\ref{Om-om}) determines the density
$j(\chi,\theta)$  which in turn gives $J(<\chi)$ and $\omega(t,\chi,\theta)$ is then
obtained from equations  (\ref{0omJ}), (\ref{1omJ}), (\ref{-1omJ}). Angular
velocity of matter, $\Omega(t,\chi,\theta)$, is then given again by equation 
(\ref{Om-om}).

If we are interested in proper azimuthal velocities, we can write
\begin{equation}
V=ar\sin\theta~\Omega,\qquad v=ar\sin\theta~\omega,
\end{equation}
and rewrite (\ref{Om-om}) as
\begin{equation}
V-v=\frac{1}{a^4(\rho+p)}\cdot\frac{j(\chi,\theta)}{r^3\sin^2\theta}.
\label{V-v}
\end{equation}
Since $|\Omega r|,|\omega r|\ll1$, we have also $|V|,|v|\ll1$. In
the case of the dust universes ($p=0$) the density obeys the conservation law $\rho a^3={\rm
constant}\equiv C$.
Equation (\ref{V-v}) then implies
\begin{equation}
V-v=\frac{j(\chi,\theta)}{Cr^3\sin^2\theta}\cdot\frac{1}{a}.
\end{equation}
This is not valid near $t\sim 0$ when $a\rightarrow 0$ due to our approximation.
For $a\rightarrow \infty$, $V-v\rightarrow 0$ --- the dragging becomes perfect.


\begin{acknowledgements}
This work started during our meeting at the Institute of Theoretical Physics
of the Charles University in Prague and continued during our stay at the
Albert Einstein Institute in Golm. We are grateful to these Institutes for their
support.
\end{acknowledgements}

A partial support from the grant GA\v CR 202/02/0735 of the Czech
Republic is also acknowledged.


\begin{thebibliography}{99}
\bibitem{foot1} Here the rotations of different spherical shells may be about different axis.
\bibitem{foot2} There $(3/2)$ should read $(3/2)^2$ and $(1-\mu^2)$ should be replaced by
$k(1-\mu^2)$ to cover both $k=\pm 1$ consistently.
\bibitem{foot3} In the latter there are typographical errors in that the final
$(1-\mu)^{3/2}$ should be $(1-\mu^2)^{3/2}$ and the minus after the
final equals sign should be a + because ${\rm d}J/{\rm d}\mu$ is negative since
$\mu=+1$ at the origin. Also $a^3$ is missing.
\end{thebibliography}
\end{document}